\documentclass[twocolumn,showpacs,aps,prb]{revtex4}
\usepackage{graphicx}
\begin{document}
\title{Possible Jahn-Teller effect in Si-inverse layers}
\author{S. Brener}
\author{S. V. Iordanski} 
\author{A. Kashuba}

\affiliation{L. D. Landau Institute for Theoretical Physics, Russian Academy of Sciences, 2 Kosygina st., 119334 Moscow}

\begin{abstract}

Jahn-Teller effect in bivalley Si(100) MOSFET under conditions of quantum Hall effect at integer filling factors $\nu=1,2,3$ is studied. This system is described by SU(4) hidden symmetry. At $\nu=2$ static and dynamic lattice deformation creates an easy-plane anisotropy and antiferromagnetic exchange and lifts the valley degeneracy. At $\nu=1,3$ Coulomb interaction is essential to produce weak easy-plane anisotropy. Three phases: ferromagnetic, canted antiferromagnetic and spin-singlet, have been found. Anisotropy energy of charged skyrmion excitation in every phase is found.

\end{abstract}
\maketitle

In a symmetric configuration of atoms the electronic state could be degenerate due to a high symmetry of the Hamiltonian. H.A.Jahn and E.Teller proved that such a symmetric configuration is unstable to a deformation lowering the overall symmetry of the Hamiltonian. This Jahn-Teller effect is well accounted for in molecules and crystals \cite{3tom}. In 2D electron gas (2DEG) under the conditions of quantum Hall effect the kinetic energy of electrons is quenched. Electron states become macroscopically degenerate forming Landau levels. Interaction may remove some of this degeneracy. Nevertheless at integer fillings one finds the global spin-valley degeneracy. In this work we raise a question whether a lattice deformation - Jahn-Teller effect - lifts this global valley degeneracy at integer fillings.  

Theory of 2DEG at integer $\nu$ uses \cite{bie} Hartree-Fock approximation in the limit of small parameter: $E_c/\hbar\omega_c$, where $E_c=e^2/l_H$ is the energy of the Coulomb interaction and $\omega_c$ is the frequency of the cyclotron resonance. At $\nu=1$ theory predicts a ferromagnetic ground state with degenerate uniform spin orientation. The elementary excitations are electron-hole pairs or neutral excitons, which correspond to gapless and noninteracting \cite{ll} spin-waves for vanishing momentum and vanishing effective $g$-factor. In the limit of the large exciton momentum the electron and the hole become independent charged excitations. A special topological spin texture in 2D ferromagnet called a skyrmion \cite{bp} has a unit charge and half the quasiparticle energy \cite{skkr}. In this paper we study such textures in the presence of Jahn-Teller effect.

In the bulk Si there is a six-fold valley degeneracy, which is reduced to two-fold for (100) orientation of 2D-plane due to a quantization of electron motion in quantum well \cite{ando}. Even for orientation (111) of 2D-plane with six equivalent valleys only two valleys connected by the time reversal symmetry are occupied by electrons in strong magnetic fields \cite{bi86}. In this paper we study the effects related to this valley degeneracy and show that such bivalley system has close analogy to the bilayer systems. The bilayer setup has been extensively studied theoretically at $\nu=1$ \cite{r,moon} and at $\nu=2$. \cite{ssz} The layer degeneracy of electron states can be described by a four component spinor in combined spin-layer space. In the works \cite{mac,ik} the total 2DEG Hamiltonian was subdivided into a symmetric part and a small anisotropic part which reduces the symmetry. Symmetric part is invariant under a hidden SU(4) group of transformations in the combined spin and valley Fock space which augment apparent group SU(2)$\otimes $SU(2) of separate transformations in spin and valley spaces. 

Electron annihilation operator can be expanded using one electron orbital functions:
\begin{equation} \label{wavefunction}
\psi_\alpha(\vec{\rho})=\sum_{\tau}\psi_{\alpha\tau}(\vec{r}) \chi(z)e^{i\tau Qz/2},
\end{equation}
where $\alpha$ is the spin index, $z$ is the perpendicular to 2D plane coordinate and $\vec{r}$ is in-plane coordinate vector. The index $\tau=\pm 1$ numerates two valley. Valley wave functions of an electron in quantum well of Si inverse layer: $\chi(z)\exp(\pm iQz/2)$, are normalized and almost orthogonal for smooth real $\chi(z)$ with the negligible overlap $\int\chi^2(z) \exp(iQz) dz$. $Q$ is the shortest distance between the valley minima in reciprocal space, it's approximately equal to $2/a_{Si}$, with $a_{Si}$ being the lattice constant. $\psi_{\alpha\tau} (\vec{r})$ is the wave function of in-plane motion of an electron in valley $\tau$ and it constitutes a four component spinor. Electrons in the system will strongly interact with phonons with momentum $\pm Q$, giving rise to scattering processes from one valley to another. The Jahn-Teller effect (JTE) in Si-inverse layers corresponds to displacement of silicium atoms in $z$-direction. The new equilibrium is determined from the balance of electron-phonon and elastic energies. Strong magnetic field is essential for JTE. In this paper we study this phenomenon in details. 

\section*{Hamiltonian of electron gas on Si interface}

The Hamiltonian of Si inverse layer describes electron system interacting with phonons:
\begin{equation}\label{hamil}
\hat{H}=\hat{H}_{e}+\hat{H}_{e-ph}+\hat{H}_{ph}.
\end{equation}
For narrow quantum well, 2DEG Hamiltonian in magnetic field can be expressed in terms of spinor $\psi_{\alpha\tau} (\vec{r})$: 
\begin{eqnarray}\label{ele}
\hat{H}_{e}\!\!&=&\!\int\!\! \psi^\dagger_{\alpha\tau}(\vec{r})\left( \frac{1}{2m} \left[-i\vec{\nabla}+ \vec{A}\right]^2\!\! +g\mu_B\vec{B} \vec{\sigma} \right)_{\alpha\beta} \!\!\psi_{\tau\beta}(\vec{r}) d^2\vec{r} \nonumber\\ &+&\!\!\! \frac{e^2}{2\kappa} \int\! \frac{\psi^\dagger_{\alpha\tau_1} (\vec{r}_1) \psi^\dagger_{\beta\tau_2}(\vec{r}_2) \psi_{\beta\tau_2}(\vec{r}_2) \psi_{\alpha\tau_1}(\vec{r}_1)} {|\vec{r}_1-\vec{r}_2|} d^2\vec{r}_1d^2\vec{r}_2,
\end{eqnarray}
where $g$ is the conduction band gyromagnet ratio, $\mu_B$ is the Bohr's magneton, $\vec{B}$ is the uniform magnetic field in $z$-direction. We neglect small terms with different valley indices in the electron density operator. We use the simplest Debay model for phonons with large momentum $\sim \pm Q$
\begin{equation}\label{Hphon}
\hat{H}_{ph}[u_z]=\frac{\rho_{Si}}{2}\int\left[\left(\partial_t u_z\right)^2 +c^2\left(\vec{\nabla} u_z \right)^2 \right] d^3\vec{\rho}, 
\end{equation}
where $u_z(\vec{\rho})$ is the lattice displacement polarized in $z$-direction. We consider only the valley-mixing electron-phonon interaction:
\begin{eqnarray}\label{Heph}
\hat{H}_{e-ph}=\Theta\int\psi^\dagger_{\alpha\tau_1}(\vec{r})\tau^\pm_{\tau_1\tau_2} \psi_{\alpha\tau_2}(\vec{r})\times \nonumber\\ \left(\int\chi^2(z) e^{\pm iQz}\nabla_z u_z (\vec{\rho})dz\right) d^2\vec{r},
\end{eqnarray}
with $\Theta$, $\kappa$ and $\rho_{Si}$ being the deformation potential, the dielectric constant and the density of Si. $\tau^{\mu}$ and $\sigma^\mu$, with $\mu=x$, $y$ and $z$, are the Pauli matrices in valley and spin space and $\tau^{\pm}=(\tau^x\pm i\tau^y)/2$. We choose the units where: $\hbar=1$, $e=c$, $B=1$ and the distance is measured in magnetic length $l_H=1$. $\vec{A}(\vec{\rho})= (0,Bx)$ is the vector-potential in Landau gauge.

In quantizing perpendicular magnetic field we expand the electron operator over Landau orbital states $\phi_{n,p}(\vec{r})$:
\begin{equation}\label{EigenFunc}
\hat{\psi}_{\alpha\tau}(\vec{r})=\sum_{n,p}\phi_{n,p}(\vec{r})\hat{c}_{np,\alpha\tau}, 
\end{equation} 
where $c^\dagger_{np,\alpha\tau}$ and $c_{np,\alpha\tau}$ are electron creation and annihilation operators in $\tau$ valley with spin $\alpha$. $n$ numerates Landau levels and continuous parameter $p$ in the Landau gauge specifies states inside one Landau level.  

We assume the quantum Hall effect conditions where electrons are confined to the lowest Landau level $n=0$ in the ground state. The electronic part of the Hamiltonian contains the Coulomb and the kinetic energy on the lowest Landau level:
\begin{equation}\label{symHam} 
H^{sym}=\frac{1}{2m}\sum_p c^\dagger_{p\alpha\tau}c_{p\alpha\tau}+\frac{1}{2} \int \frac{d^2\vec{q}}{(2\pi)^2} V(\vec{q}) N(\vec{q}) N(-\vec{q}), 
\end{equation} 
where $V(\vec{q})=2\pi e^2/\kappa q$ is 2D Coulomb interaction and the electron density operator is:
\begin{equation}\label{N0}
N(\vec{q})\ =\ \sum_p\ c^\dagger_{p\alpha\tau} c_{p-q_y\,\alpha\tau} \exp\left[-iq_x(p-q_y/2) -q^2/4\right].
\end{equation} 
Unitary transformations of electron operators $c_{\alpha\tau_1,p}= U_{\alpha\tau_1,\beta\tau_2} c_{p\beta\tau_2}$ in the combined spin and valley space leave the Hamiltonian (\ref{symHam}) invariant, where $U$ is a matrix from SU(4) Lee group. For Landau level filling factor $\nu=1,2,3$ we assume that the ground state is uniform over $p$-orbitals with electrons of spin $\alpha_i$ and valley $\tau_i$ fill every orbital $p$ of the lowest Landau Level:
\begin{equation}\label{PsiRitor}
\Psi(\alpha_i\tau_i) =\ \prod_{i=1}^{\nu}\prod_p c^\dagger_{p\alpha_i\tau_i}\ |vac\rangle.
\end{equation}
One can easily check by inspection that any such wave-function (\ref{PsiRitor}) represents an eigen function of the $H^{sym}$ (\ref{symHam}). The state (\ref{PsiRitor}) is degenerate and a set of related eigen states can be generated by applying uniform rotations $U$. 

The remaining terms in the Hamiltonian (\ref{hamil}) consist from a valley splitting term due to a singularity of the well potential on Si/SiO$_2$ interface \cite{ando}, Zeeman term, electron-phonon interaction and the phonon energy:
\begin{eqnarray}\label{anisHam}
H^{an}\ =\ -t\sum_p c^\dagger_{\alpha\tau_1p} \tau^x_{\tau_1\tau_2} c_{\alpha\tau_2p} - \nonumber\\ -|g|\mu_BH \sum_p c^\dagger_{\alpha\tau p} \sigma^z_{\alpha\beta} c_{\beta\tau p}+H_{e-ph}+H_{ph}  ,      
\end{eqnarray} 
where $t$ is the phenomenological valley-splitting constant. This part breaks down the SU(4) symmetry and lifts degeneracy of the eigen states of SU(4)-symmetric Hamiltonian (\ref{hamil}). The splitting of energy levels is determined by matrix elements of weak anisotropy Hamiltonian (\ref{anisHam}) projected onto a linear space of the symmetric Hamiltonian (\ref{symHam}) level degeneracy. There is no renormalizations of the anisotropy Hamiltonian parameters (\ref{anisHam}) due to electron-electron interaction in symmetric part (\ref{symHam}). Thus the Hartree-Fock approach is valid up to $O(H^{an}/\omega_H)$ terms for the $\nu=1,2,3$ cases.

The total Hamiltonian (\ref{symHam},\ref{anisHam}) can be treated as in the theory of magnetism. There exists a local order parameter matrix $\hat{Q}(\vec{r})$ and the total bivalley energy can be expressed in terms of this order parameter. The energy of isotropic part (\ref{symHam}) will be expanded in powers of spatial variations of order parameter $\vec{\nabla} \hat{Q}(\vec{r})$ due to Goldstone theorem, whereas in the anisotropy parts we retain only homogeneous order parameter $\hat{Q}$.

\section*{SU(4) Symmetric Case}

Non-homogeneous state can be generated from the ground state (\ref{PsiRitor}) by slow rotations, with the effective action being dependent on rotation matrix $U(t,\vec{r})$:
\begin{equation}\label{ActLagr}
S[U]=-i\ {\rm Tr}\ \log\int{\cal D} \psi{\cal D}\psi^\dagger\ \exp \left(i\int{\cal L} dt\right) 
\end{equation} 
where the spin-valley symmetric Lagrangian of 2DEG is:
\begin{eqnarray}\label{GradHam} 
&&{\cal L}=\!\int \psi^\dagger_\alpha\left[ i{\partial\over\partial t}- \frac{1}{2m} \left(-i\vec{\nabla}+\vec{A}_0+ \vec{\Omega}\right)^2 \right]_{\alpha\beta} \psi_\beta\,d^2\vec{r} \nonumber\\ &&-\int \psi^\dagger_\alpha \Omega^t_{\alpha\beta} \psi_\beta\ d^2\vec{r}+ \frac{1}{2}\sum_{\vec{q}} V(\vec{q}) N(\vec{q})N(-\vec{q}),
\end{eqnarray} 
where $ \Omega^t=-i\,U^\dagger \partial_t U$, $\vec{\Omega}= -i\, U^\dagger\vec{\nabla}U$. We treat nonhomogeneous matrix $U$ as a classical filed. Therefore the effective action (\ref{ActLagr}) describes a macroscopic motion of the corresponding spin-texture. The essential points here is the slow rotation of $U$ on the microscopic scale given by magnetic length and the use of gradient expansion assuming locally ferromagnetic ground state. The effective action (\ref{ActLagr}) for multivalley was found in Ref.\cite{akl}. We derive it following the method of Ref.\cite{IPF99}.

We consider three filling factors of uniform ground state (\ref{PsiRitor}): $\nu=1$ with $\Psi(\uparrow+1)$, $\nu=2$ with $\Psi(\uparrow+1,\downarrow-1)$ and $\nu=3$ reduces to the case $\nu=1$ under the electron-hole transformation. Reference state is alternatively given by an electron occupation matrix: $N_{\alpha\tau_1,\beta\tau_2}= \left(1,0,0,0\right)$, in the case $\nu=1$, and $N_{\alpha\tau_1,\beta\tau_2}= \left(1,1,0,0 \right)$, in the case $\nu=2$. Electron Green function on the lowest Landau level for the reference ground state of Hamiltonian (\ref{GradHam}) with $\vec{\Omega}=0$ reads:
\begin{equation}\label{GreenF0} 
G^0_{\alpha\tau_1,\beta\tau_2}(\epsilon,p)=\frac{N_{\alpha\tau_1,\beta\tau_2}}{\epsilon+E_0+\mu-i0}+\frac{(\hat{1}-N)_{\alpha\tau_1,\beta\tau_2}} {\epsilon+\mu+i0},
\end{equation} 
where $\mu=-E_0/2$ is the chemical potential and exchange constants in the limit of vanishing thickness of electron layer are:
\begin{equation}\label{E0E1} E_0=2E_1=\sqrt{\pi\over 2}{e^2\over\kappa l_H}. 
\end{equation}

\begin{figure}
\includegraphics{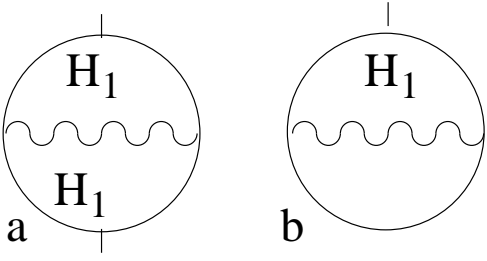}
\caption{\label{Diagrams} Second order a) and the anomalous first order b) diagrams. Solid and wavy line represent the electron Green functions and Coulomb interaction.}
\end{figure} 

The effective action is $S[\Omega]=S_0[\Omega]+S_2[\Omega]$, where $S_0[\Omega]=i\,{\rm Tr}\log(G/G_0)$ whereas $S_2[\Omega]$ is represented by two diagrams on Fig.\ref{Diagrams}. The first order perturbation correction to the Green function: $\delta G=G-G_0$, contains an electron propagation on the first excited Landau Level \cite{IPF99} with the Green function:
\begin{equation}\label{GreenF1} 
G^1_{\alpha\tau_1,\beta\tau_2} (\epsilon)=\frac{N_{\alpha\tau_1,\beta\tau_2}} {\epsilon-\omega_c +E_1+\mu+i0}+\frac{(\hat{1}-N)_{\alpha\tau_1, \beta\tau_2}} {\epsilon-\omega_c+\mu+i0}. 
\end{equation} 

The two terms of the Hamiltonian depend on the gradient matrix field:
\begin{equation}\label{H1} 
H_1=\frac{1}{2m}\int \psi^\dagger_{\alpha\tau_1}\left( \Omega^+\hat{\Pi}_- +\hat{\Pi}_+\Omega^-\right)_{\alpha\beta\tau_1\tau_2} \psi_{\beta\tau_2}\ d^2\vec{r}, 
\end{equation}
\begin{equation}\label{H2} H_2=\frac{1}{2m}\int \psi^\dagger_{\alpha\tau_1} \left(\vec{\Omega}^2- \epsilon_{\mu\nu}\partial_\mu \Omega_\nu\right)_{\alpha\beta\tau_1\tau_2} \psi_{\beta\tau_2}\ d^2\vec{r}, 
\end{equation} 
where $\Omega_\pm=\Omega_y\mp i\Omega_x$ ($\Omega_-= \Omega_+^*$) and the differential operators $\hat{\Pi}_\pm$ shift an electron between the adjacent Landau Levels: $\hat{\Pi}_- \phi_{np}(\vec{r}) =\sqrt{2n}\phi_{n-1p} (\vec{r})$, $\hat{\Pi}_+\phi_{n-1p} (\vec{r})=\sqrt{2n} \phi_{np}(\vec{r})$, though only the $n=0$ and $n=1$ Landau Level states are relevant for our problem. An expansion of the 2DEG action up to the second power of the Hamiltonian (\ref{H1}) gives:
\begin{equation}\label{ActExp0}
S_0[\Omega]=i{\rm Tr}\left(H_1G_0\right)+ \frac{i}{2}{\rm Tr}\left(H_1G_0H_1G_0\right)+ i{\rm Tr}\left(H_2G_0\right). 
\end{equation} 
The Hartree-Fock diagram on Fig.\ref{Diagrams}a has been calculated in Ref.\cite{IPF99} whereas the anomalous diagram on Fig.\ref{Diagrams}b is calculated in the Appendix A. The resulting effective Lagrangian is:
\begin{eqnarray}\label{LG} 
&& {\cal L}_{eff}[\Omega]= \int \Big({\rm Tr}(\Omega_t(t,\vec{r})\hat{N}) + \frac{E_0+E_1}{2} \epsilon_{\mu\nu} \nabla_\mu\Omega^z_\nu -\nonumber\\ &-& \frac{E_1}{2} {\rm Tr} \left[N \Omega_+(t,\vec{r}) (\hat{1}-N)\Omega_-(t,\vec{r})\right] \Big) \frac{d^2\vec{r}}{2\pi},
\end{eqnarray} 
where $ \Omega^z_\mu=-i\,{\rm Tr}\left(N U^\dagger \partial_\mu U\right)$. The matrices $N$ and $\hat{1}-N$ are projector operators onto the physical rotations that form a sub-set of SU(4) Lee group. The matrix field $\Omega_\mu$ can be expanded in the basis of fifteen generators of SU(4) group: $\{\Gamma^l\}$, with $l=1..15$. They form two complementary sets: the first $even$ set includes those generators that do commute with $N$, whereas the second $odd$ set includes the remaining generators. Generators of the $even$ set constitute an algebra itself. This algebra has a normal Abelian subalgebra formed by a single traceless generator: $N-\nu/4$. A Lee group built around the $even$ set of generators is called a stabilizer sub-group ${\cal S}$ of SU(4) group. The $odd$ set always contains an even number of generators: eight in the case of $\nu=2$ and six in the case $\nu=1,3$.

The Hamiltonian is invariant under the time reversal symmetry. The time reversal operation transforms rotation matrix as $U\rightarrow U^*$, gradient field as $\vec{\Omega} \rightarrow -\vec{\Omega}^T$ and also inverts the magnetic field $B^z$. Accordingly we rewrite the energy in Lagrangian (\ref{LG}) in time reversal symmetric way:
\begin{eqnarray}\label{GoldHam} 
E_{eff}[\Omega] &=& \int{d^2\vec{r}\over 2\pi}\Big[ \frac{E_1}{2}{\rm Tr} \left(N \Omega_\mu (\hat{1}-N) \Omega_\mu\right) - \nonumber\\  &-& \frac{E_0}{2}\ {\rm sgn}(B^z) \epsilon_{\mu\nu} \nabla_\mu \Omega_\nu \Big],
\end{eqnarray}
where we used identity $\epsilon_{\mu\nu} {\rm Tr}\left( \Omega_\mu \Omega_\nu N\right)= i \epsilon_{\mu\nu} \nabla_\mu\Omega^z_\nu$. It follows immediately that $E_{eff}\ge 0$. The first term is the gradient energy whereas the second term is proportional to the topological index of the spin texture:
\begin{equation}\label{Index}
{\cal Q}=\int\epsilon_{\mu\nu}\nabla_\mu\Omega^z_\nu\ \frac{d^2\vec{r}}{2\pi}= Z, 
\end{equation} 
where $Z$ is integer. The states with $Z\neq 0$ are called skyrmions in the name of T.H.R.Skyrme who first considered such textures. The case ${\cal Q}=\pm 1$ corresponds to the simplest spin skyrmions in the first valley which can be rotated by a SU(4) matrix to become a general bivalley skyrmion. The spin stiffness in $E_{eff}$ (\ref{GoldHam}) coincide identically with that of the one-valley case \cite{IPF99}, which means that the bivalley skyrmion energy is the same as found for one valley. Also the charge in the bivalley skyrmion core is distributed over the two valleys with long divergent tails: $n_{\pm 1}(\vec{r})\sim\pm 1/\left(R^2+r^2\right)$, if one neglects anisotropy. But the total charge of the two valleys follows a convergent distribution identical with a charge density in the one valley case \cite{skkr}:
\begin{equation}\label{ChargeDens1}
n(\vec{r})=\frac{\epsilon_{\mu\nu}\nabla_\mu\Omega^z_\nu(\vec{r})}{2\pi}= \frac{R^2}{\pi \left(R^2+r^2\right)^2}, 
\end{equation}
where $R$ is the radius of the skyrmion core. 

Expectation value of any operator $A$ in the ground state $\langle A\rangle={\rm Tr}(A\hat{Q})$ can be expressed in terms of the order parameter matrix:
\begin{equation}\label{OrderQ}
\hat{Q}(\vec{r})=U(\vec{r})NU^+(\vec{r}). 
\end{equation} 
Obviously rotations from the little sub-group ${\cal S}$ leave the order parameter intact. Thus, rotations in (\ref{OrderQ}) can be restricted to a physical space of the bivalley 2DEG - complex Grassmannian manifold: $U(4)/U(\nu)\otimes U(4-\nu)$. We rewrite (\ref{GoldHam}) in terms of $\hat{Q}$:
\begin{equation}\label{SigmaModel} 
E_{eff}[\hat{Q}]={E_1\over 4}\int {\rm Tr}\left(\vec{\nabla} \hat{Q}\vec{\nabla}\hat{Q}\right) \frac{d^2\vec{r}}{2\pi}-\frac{E_0}{2}{\cal Q}. 
\end{equation} 
In this representation the topological index ${\cal Q}$ appears as an index of a map of 2D plane on the coset space of the order parameter. The rule (\ref{Index}) is a consequence of a well known homotopy group identity:
\begin{equation}\label{Pi2Coset} 
\pi_2\left({U(4)\over U(\nu)\otimes U(4-\nu)}\right)=Z. 
\end{equation} 
This identity is proven in Appendix B.

Varying the effective Lagrangian consisting of the first kinetic term in (\ref{LG}) (rewritten as the Wess-Zumino term \cite{akl}) and the energy (\ref{SigmaModel}) we find the matrix Landau-Lifshitz equation:
\begin{equation} \label{LLQ}
\left[\hat{Q}\ \partial_t\hat{Q}\right]_-= \frac{E_1}{2}\vec{\nabla}^2\hat{Q}.
\end{equation}
It describes three ($\nu=1$) and four ($\nu=2$) degenerate 'spin-valley'-waves with dispersion $\omega(\vec{q})=E_1 q^2/2$. 

\section*{Jahn-Teller effect}

The Jahn-Teller effect is the deformation of a lattice lowering $H_{e-ph}+H_{ph}$ energy. Phonons interact with the local electron density. We illustrate construction of electron density in the case $\nu=1$ with fixed spin of electron. Electron annihilation operator (\ref{wavefunction}) is expanded in basis of two mutually orthogonal valley wave function $\tau=\pm 1$. The density operator $\hat{n}(\vec{\rho})= \hat{\psi}^\dagger (\vec{\rho}) \hat{\psi} (\vec{\rho})$ has a diagonal element uniform over 2D plane: $\chi^2(z)\langle \psi^\dagger_{\tau} (\vec{r}) \psi_{\tau}(\vec{r}) \rangle=\nu\chi^2(z)/2\pi l_H^2$, where average is taken over the ground state (\ref{PsiRitor}), and the term oscillating in $z$ direction: $\delta\hat{n}(\vec{\rho})= \chi^2(z)(\psi^\dagger_{+1} (\vec{r}) \psi_{-1}(\vec{r}) e^{iQz}+h.c.)$. We use electronic order parameter (\ref{OrderQ}) to express the average oscillating density as:
\begin{equation}
\delta n(\vec{\rho})=\frac{\chi^2(z)}{2\pi l_H^2}\left( {\rm Tr}(\hat{Q}\tau^+) e^{iQz}+ {\rm Tr}(\hat{Q}\tau^-) e^{-iQz}\right)
\end{equation}
Diagonal density term couples to phonons with in-plane momentum $\vec{q}_{\perp}=0$ and can be neglected in thermodynamic limit. Oscillatory part of the average electron density allows for lowering the energy by static lattice deformation also oscillating with wavevector $Q$ inside the quantum well:
\begin{equation}\label{ust}
\nabla_z u^z_{st}(\vec{\rho})=\frac{\Theta}{\rho_{Si}c^2}\delta n(\vec{\rho}).
\end{equation}
This deformation is a phonon 'condensate' or the Jahn Teller effect. The corresponding energy gain is found from the minimization of $H_{e-ph}+H_{ph}$ as:
\begin{equation}\label{static}
E_{st}=-{\cal N} G {\rm Tr}(\hat{Q}\tau^+) {\rm Tr}(\hat{Q}\tau^-).
\end{equation} 
where ${\cal N}=AeB/2\pi \hbar c$ is the number degeneracy of the Landau level and $G= \Theta^2\int\chi^4(z)dz\ /2\pi l_H^2 \rho_{Si} c^2$ is the strength of electron-phonon interaction. Integral is the inverse width of the 2D layer in $z$-direction.

The static deformation gives new equilibrium position of the lattice. Assuming static deformation small we neglect the change of phonon spectrum due to anharmonic effects. But we take into account the dynamical phonons which produce a polaronic effect in the second order of electron-phonon interaction. Expanding the phonon field around new equilibrium: $u^z(\vec{\rho})= u^z_{st}(\vec{\rho}) +\delta u^z(\vec{\rho})$, we obtain:
\begin{eqnarray}\label{e-ph}
\hat{H}_{e-ph}&=&\Theta\int\psi^\dagger_{\alpha\tau_1}(\vec{r})\tau^\pm_{\tau_1\tau_2} \psi_{\alpha\tau_2}(\vec{r})\times \nonumber\\ && \left(\int\chi^2(z) e^{\pm iQz}\nabla_z \delta u_z (\vec{\rho})\ dz\right) d^2\vec{r},
\end{eqnarray}
where the lattice deformation $$\delta u_z(\vec{\rho}) =\sum_{\vec{q}} \sqrt{\frac{\hbar}{2V\rho_{Si}\omega(q)}} \left(b_{\vec{q}} e^{i\vec{q}\vec{\rho}}+ b_{\vec{q}}^\dagger e^{-i\vec{q}\vec{\rho}}\right),$$ can be expanded in phonon creation and annihilation operators $b^\dagger_{\vec{q}}$ and $b_{\vec{q}}$. In order to find the ground state energy we sum up diagrams for the thermodynamic potential expanded in powers of (assumed to be) weak electron-phonon and Coulomb interactions using the Matsubara method \cite{agd}. We express the energy in terms of the order parameter matrix $\hat{Q}$.

Electron's Green function on the lowest Landau level can be expressed in terms of the order parameter $\hat{Q}$:
\begin{equation}\label{green}
G^0(\varepsilon_n,p)=\frac{\hat{Q}}{i\varepsilon_n+E_0/2}+\frac{\hat{1}-\hat{Q}} {i\varepsilon_n-E_0/2}.
\end{equation}
We attribute the coordinate dependent electron wave-functions to the interaction vertices. The phonon's Green's function is \cite{agd}:
\begin{equation} \label{phonon}
D(\omega_n,q)=-\frac{\omega^2(q)}{\omega_n^2+\omega(q)^2}.
\end{equation}

At small transfered momentum the Coulomb line is much larger then the phonon propagator. In the opposite case of large transfered momentum with $\pm Q$ $z$-component we neglect the Coulomb interaction. Therefore in the Coulomb vertex the valley index is conserved as well as the spin index. Due to condition $\hat{Q}(1-\hat{Q})=0$ all one loop diagrams with Coulomb lines only vanish. 

\begin{figure}
\includegraphics{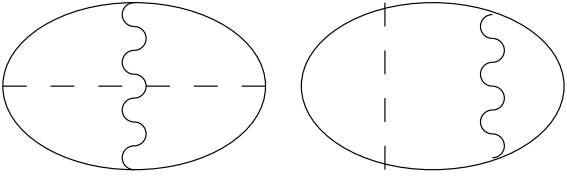}
\caption{\label{exchange} Two diagrams that gives easy-plane anisotropy in the case $\nu=1$. Solid, dashed and wavy lines are the electron, phonon propagation and the Coulomb interaction}
\end{figure} 

The expression for the electron-phonon vertex in Landau gauge reads: $$g(pp',\vec{q})= \frac{\Theta}{\sqrt{\rho_{Si}c^2}} \delta_{p,q_y+p'} e^{-q^2_{\perp}/4-q_x(p+p')/2} \chi^2{q_z-Q},$$ where $\vec{q}$ is the phonon momentum and the envelope function $\chi^2(q_z)= \int\chi^2(z) e^{iq_zz}\,dz$ has characteristic width $l^{-1}$ in momentum space. The polaron contribution to the energy assuming weak Coulomb interaction: $E_0/\omega(Q)\to 0$, is:
\begin{eqnarray}\label{pol}
&& E^0_{e-ph}=\sum_{\varepsilon_n\omega_npp'}{\rm Tr}\left(G^0(\varepsilon_n,p)\tau^+ G^0(\varepsilon_n+\omega_n,p')\tau^-\right)  \nonumber\\ &&\sum_{\vec{q}} g^2(pp',\vec{q}) D(\omega_n,q) = {\cal N} G \left({\rm Tr}(\hat{Q}\tau^+\hat{Q}\tau^-)-\frac{\nu}{2}\right).
\end{eqnarray}
Summing it up with (\ref{static}) we get the total electron-phonon energy in the second order of electron-phonon interaction. Here we neglect small terms of the order $(Ql)^{-4}$ due to derivative jump of wave  function $\chi(z)$ at interface. \cite{r} At $\nu=1$ the energy (\ref{static},\ref{pol}) is isotropic in spin-valley space as the two expressions are canceled exactly. If we neglect the Coulomb interaction corrections to electron-phonon interaction we find that the degeneracy in isospin direction is not lifted. Physically, polaron energy is a single electron energy and there is no anisotropy operator for a single electron because of the Pauli matrix identity: $\tau_i^2=1$.

Coulomb interaction creates the easy-plane anisotropy in the case $\nu=1$ and the essential diagrams are shown on Fig.(\ref{exchange}). The dashed line represents the 2D Coulomb potential $V(q)$ whereas the solid and wavy lines represent the electron (\ref{green}) and phonon (\ref{phonon}) propagators. Direct calculation neglecting the dependence of $\omega(Q)$ on in-plane momentum $q_{\perp}$ shows that the sum of two diagram on Fig.(\ref{exchange}) is
\begin{equation}
E^\delta_{e-ph}=-{\cal N}G\delta {\rm Tr}\left(\hat{Q}\tau^+\hat{Q}\tau^-\right),
\end{equation}
where $ \delta=E_1/\omega(Q)$ is the ratio of the Coulomb exchange energy:
\begin{equation}\label{coulombconstants}
E_1= \frac{e^2}{\pi\kappa} \int\frac{d^2q_{\perp}\, dq_z}{q_{\perp}^2+q^2_z}|\chi^2(q_z)|^2 e^{-q_{\perp}^2l_H^2/2}\left(1-e^{-q_{\perp}^2l_H^2/2}\right),
\end{equation}
and the energy of valley-mixing phonon. Obviously $\delta>0$ ($E_1>0$) and the anisotropy is the easy-plane. The above result is valid when $E_0\ll\omega(Q)$. Otherwise we expect this result to be valid qualitatively. The constant (\ref{coulombconstants}) in the limit $l\ll l_H$ reduces to (\ref{E0E1}).

The sum of polaron (\ref{static},\ref{pol}) and the next order $E^\delta_{e-ph}$ energies gives the easy-plane anisotropic energy:
\begin{eqnarray}\label{lp}
E_{ep} &=& G\int\big[-{\rm Tr}(\hat{Q}\tau^+) {\rm Tr}(\hat{Q}\tau^-)+\nonumber\\ && +(1-\delta){\rm Tr}\left(\hat{Q}\tau^+\hat{Q}\tau^-\right) \big]\frac{d^2\vec{r}}{2\pi}.
\end{eqnarray}
We conclude that the JTE is energetically favorable and it gives rise to the easy-plane valley anisotropy (\ref{lp}).

The easy-plane anisotropy changes dispersion of collective excitations. Let us consider this effect in the limit of vanishing valley symmetry breaking $t$ and Zeeman $h$ anisotropies. We add the easy-plane term (\ref{lp}) into the Landau-Lifshitz equation (\ref{LLQ}):
\begin{equation} \label{LLQ1}
\left[\hat{Q}\ \partial_t\hat{Q}\right]_-= \frac{E_1}{2}\vec{\nabla}^2\hat{Q}+ G\delta\left(\tau^+{\rm Tr}(\hat{Q}\tau^-)+\tau^-{\rm Tr}(\hat{Q}\tau^+)  \right)
\end{equation}
This equation can be linearized in the vicinity of $Q=N$ and describes two 'spin'-modes (unchanged from the symmetric case) and one acoustic mode with dispersion $\omega=\sqrt{G\delta E_1}q$ in analogy to the bilayer case \cite{r}. 

\section*{Phase diagram}

The anisotropic bivalley energy in the uniform state is a diagonal matrix element of the anisotropic Hamiltonian (\ref{hamil}) expressed in terms of the order parameter $\hat{Q}$:
\begin{eqnarray}\label{anisotropy}
E_{an}/{\cal N}=-t\ {\rm Tr}{(\hat{Q}\tau^x)}- h\ {\rm Tr}{(\hat{Q}\sigma^x)}+ \nonumber\\ G\left[(1-\delta){\rm Tr}\left(\hat{Q} \tau^+\hat{Q}\tau^-\right)-{\rm Tr}(\hat{Q}\tau^+) {\rm Tr}(\hat{Q}\tau^-)\right],
\end{eqnarray}
where $h=|g_e|\mu_BH$. The order parameter $\hat{Q}$ can be parameterized by six ($\nu=1,3$) or eight  ($\nu=2$) angles. Diagonal matrix elements are real despite the fact that in an external magnetic field there is no time reversal symmetry. Therefore the ground state can be chosen real, produced from the reference state by SO(4) sub-group rotations. This sub-group has 6 parameters with two of them falling into the denominator sub-group ($\nu=2$). One of the remaining four angles corresponds to global rotation of all spins and is fixed by the magnetic field direction $z$. Thus, the ground state differs from the reference state by just three rotations ($\nu=2$). 

First we consider the case $\nu=1$. The ground state is $U\Psi(\uparrow+1)$, where $U=P(\vartheta)R(\theta)$. Matrix $R$ rotates the $\uparrow$-spin component in valley space and then matrix $P$ rotates the $\pm$-valley components of the resulting state in spin space by angle $\pm\vartheta$. Substituting $\hat{Q} (\vartheta,\theta)$ into (\ref{anisotropy}) we obtain the energy per one electron:
\begin{equation} \label{nuone} 
E_{an}^{1}=-h\sin{\vartheta}-t\sin{\theta}+ G\delta\cos^2{\theta}. 
\end{equation}
The minimum of this energy is reached at $\vartheta=\pi/2$ and $\theta=\pi/2$ corresponding to the ferromagnet phase in spin and valley spaces. The case $\nu=3$ is identical to $\nu=1$ due to electron-hole symmetry. 

\begin{figure}
\includegraphics{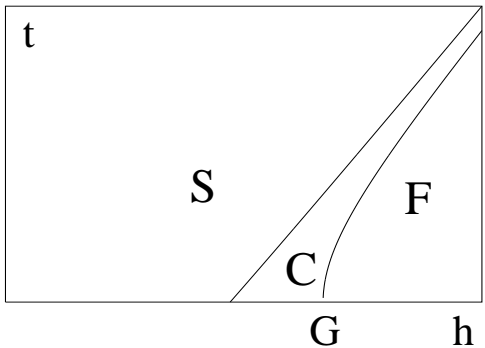}
\caption{\label{dia} Phase diagram in the $\nu=2$ $\delta=0$ case. F,S,C is ferromagnetic, spin-singlet and canted antiferromagnetic phases.} 
\end{figure}

In the case $\nu=2$ the ground state is $U\Psi(\uparrow +1,\downarrow -1)$ where $U=P(\vartheta) R(\theta_\uparrow, \theta_\downarrow)$. Matrix $R$ rotates the $\uparrow,\downarrow$-spin components by angles $\theta_{\uparrow,\downarrow}$ in the valley space and then matrix $P$ rotates the $\pm 1$-valley components by angles $\pm\vartheta$ in spin space. We find $\theta_\uparrow= \theta_\downarrow$. Substituting $\hat{Q} (\vartheta,\theta)$ into (\ref{anisotropy}) we find for $\delta=0$:
\begin{eqnarray}\label{nutwo} 
E_{an}^{2}=-2t\cos{\vartheta}\sin{\theta}-2h\sin{\vartheta}\cos{\theta}+\nonumber\\
+G(1-\delta)\left(\frac{1}{2}\cos^2{\theta}-\cos^2{\vartheta}\right)-G\delta\cos^2{\vartheta} \sin^2{\theta}.
\end{eqnarray}
The last term represent easy-plane anisotropy and antiferromagnetic exchange interaction between spins in two valleys. The phase diagram in the case $\delta=0$ is shown on Fig.\ref{dia}. Here $F$ is the spin ferromagnetic and the valley singlet phase ($\vartheta=\pi/2$, $\theta=0$), $S$ is the spin singlet and the valley ferromagnetic phase ($\vartheta=0$, $\theta=\pi/2$), $C$ is the canted antiferromagnetic phase in spin and valley spaces. 

At $\delta=0$ we find two lines of continuous phase transition between F and C phases: $(h-G)(h-G/2)=t^2$, and between S and C phases: $(t+G)(t+G/2)=h^2$. Small positive $\delta$ transforms the S-C phase transition into a discontinuous first order one. C phase ends at some critical $h_c(\delta)$ and the direct S-F first order transition occurs at $h>h_c(\delta)$. For $\delta>0.33$ C phase disappears and there is a line of direct first order S-F transition: $t=h-G(3-\delta)/4$.

At typical experimental magnetic field $B\sim 5-10T$ Zeeman energy in Si is few times greater then valley splitting $t\sim 2K$ \cite{pud}. To our knowledge the deformation potential at large phonon momentum is unknown. Therefore we could only guess the electron-phonon energy $G\sim 0.2-2K$. Probably to observe phases of Fig.\ref{dia} in Si one would need to artificially lower the effective electron $g$-factor. 

\section*{Anisotropic energy of skyrmion}

Many experiments found activation gap for diagonal conductivity in QHE systems. For symmetry broken 'ferromagnet' systems theory predicts that conductivity is mediated by charged topological textures - skyrmions, with activation energy being determined by large exchange constant $E_1$ (\ref{SigmaModel}). Contrary to this experiments find much lower gaps \cite{shukla}. One possible explanation is that due to intrinsic inhomogeneity - defects or long range slow variation of electrostatic potential - skyrmions of opposite topological charges are already present in the system and experiments measure either depinning activation energy or skyrmion mobility activation energy. In both cases gap is determined by the small anisotropic part of skyrmion energy. In this section we find this anisotropic energy for a bivalley skyrmion.

Belavin and Polyakov found the skyrmion solution in the case of $S^2$ order parameter \cite{bp}. Their solution is readily generalized for general Grasmanian order parameter. The non-homogeneous order parameter that represents one skyrmion in symmetric case with $|{\cal Q}|=1$ is given by:
\begin{equation} \label{SkyRotat} 
\hat{Q}_{BP}(z)=\hat{N}+ \frac{1}{R^2+|z|^2} \left( \begin{array}{cc} -R^2 |v_f\rangle\langle v_f| & zR |v_f\rangle\langle v_e|\\  \bar{z}R |v_e\rangle\langle v_f|& R^2 |v_e\rangle\langle v_e| \end{array}  \right)       
\end{equation}
where $z=x+iy$, $R$ is the radius of skyrmion core and $|v_f\rangle$ and $|v_e\rangle$ are the two vectors of size $\nu$ and $4-\nu$. We choose $|v_{fe}\rangle=(1,0..0)$ in (\ref{SkyRotat}). The skyrmion order parameter has to be rotated by a homogeneous matrix $U$ calculated in the previous section in such a way that the order parameter far away from the skyrmion center minimizes the anisotropy energy. In addition to this rotation we have to allow the global rotations $W$ from the denominator sub-group ${\cal S}$ that transform the skyrmion order parameter (\ref{SkyRotat}): $\hat{Q}(\vec{r})=UW\hat{Q}_{BP}(z)W^+U^+$. For the case $\nu=2$ the matrix $W$ can be parameterized by seven angles:
\begin{equation}\label{Wdenom} 
W=\left( \begin{array}{cc} \displaystyle W_f & 0 \\ \displaystyle 0 & W_e \end{array} \right) 
\end{equation} 
where
\begin{equation} 
W_{ef}=\left( \begin{array}{cc} \displaystyle \cos\frac{\beta_{ef}}{2} e^{i(\gamma_{ef}+\alpha_{ef})} & \sin\frac{\beta_{ef}}{2} e^{i(\gamma_{ef}-\alpha_{ef})} \\ \displaystyle -\sin\frac{\beta_{ef}}{2} e^{i(-\gamma_{ef}+\alpha_{ef})} & \cos\frac{\beta_{ef}}{2} e^{i(-\gamma_{ef}-\alpha_{ef})} \end{array} \right).
\end{equation} 
The additional seventh parameter angle of the denominator sub-group rotates the coordinates: $z\rightarrow e^{i\gamma_7}z$. We find explicitly that the skyrmion anisotropic energy does not depend on phases $\gamma_e$, $\gamma_f$ and $\gamma_7$ whereas $\alpha_e=0$ and $\alpha_f=\pi$ correspond to the energy minimum. Important are two angles $\beta_e$ and $\beta_f$ which rotate the order parameter inside the core of skyrmion. Besides these angles the skyrmion energy depends on conformal parameters of the BP solution. We consider here only the simplest case of skyrmion topological index: $|{\cal Q}|=1$ (\ref{SkyRotat}) where only one such parameter is $R$. There are different spatial integrals that we encounter in the calculation. One integral is logarithmically large:
\begin{equation}\label{Acons}  
K(R)=\left({R\over l_H}\right)^2\log\left(\frac{l_H}{R}\sqrt{\frac{E_1}{{\cal E}^{sk}_{min}}}\right),
\end{equation}
and we neglect all others. In this way we find Zeeman energy of skyrmion: $E^{sk}_Z=K(R){\cal Z}$, where
\begin{equation}\label{Zeeman}
{\cal Z}=2h\left(2\sin\vartheta\cos\theta- \cos\vartheta(\sin\beta_f+ \sin\beta_e) \sin\theta\right)
\end{equation} 
The bare valley splitting energy of skyrmion reads: $E^{sk}_t= K(R){\cal T}$, where
\begin{equation}\label{Hopping} 
{\cal T}=2t\left(2\cos\vartheta\sin\theta\ - \sin\vartheta(\sin\beta_f+\sin\beta_e)\cos\theta \right)
\end{equation}
And finally the Jahn-Teller energy of skyrmion reads: $G_{sk}= K(R){\cal G}$, where
\begin{eqnarray}\label{CouSky} 
{\cal G} &=& G\Big[-\frac{1}{2}+3\cos^2\vartheta +\frac{1}{2}\cos^2\theta \cos(\beta_e+\beta_f) - \nonumber\\  && - \frac{3}{2}\cos^2\theta- \left(\frac{1}{2}-2\cos^2\vartheta\right) \cos(\beta_e-\beta_f)\Big].
\end{eqnarray} 
The total energy of a skyrmion includes also a direct Coulomb energy due to repulsion of an additional charge distribution $n(r)$ inside the skyrmion core (\ref{ChargeDens1}):
\begin{equation}\label{CouCharge}
E^{sk}_C=\int\frac{e^2n(\vec{r})n(\vec{r'})}{2\kappa|\vec{r} -\vec{r'}|}d^2\vec{r}d^2\vec{r'} =\frac{3{\pi}^2}{64} \frac{e^2}{\kappa R}={\cal E}_C\frac{l_H}{R}.
\end{equation}

\begin{figure}
\includegraphics{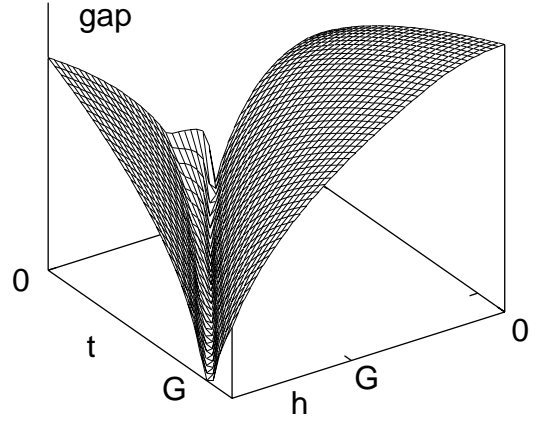}
\caption{\label{angap} Anisotropic part of skyrmion energy gap at $\nu=2$ and $\delta=0$}
\end{figure}

The minimum of the total anisotropic skyrmion energy (\ref{Zeeman},\ref{Hopping},\ref{CouSky}): ${\cal E}^{sk}={\cal Z}+{\cal T}+{\cal G}$, over the two parameters $\beta_e$ and $\beta_f$ was found numerically and is denoted as: ${\cal E}^{sk}_{min}$. Next, we find a minimum of the total skyrmion energy including (\ref{CouCharge}): $E^{sk}=K(R){\cal E}^{sk}_{min}+{\cal E}_C/R$, with respect to the skyrmion radius $R$:
\begin{equation}\label{GapSk} 
\Delta=\frac{E_1|{\cal Q}|-E_0{\cal Q}}{2}+\frac{3}{2}\left({\cal E}^{sk}_{min} {\cal E}_C^2 \log\frac{E_1}{{\cal E}_{min}^{sk}}\right)^{1/3}. 
\end{equation} 
This formula is valid in the limit ${\cal E}^{sk}_{min}\ll E_1$. The resulting anisotropic part of a skyrmion gap is shown on the Fig.\ref{angap}. Note that the prominent minimum of the skyrmion anisotropy energy gap coincide with the phase transition line C-F from the phase diagram Fig.\ref{dia}. 

This skyrmion anisotropic energy is similar to the bilayer case \cite{ik} with $t$ being the hopping constant between the two layers. In the experiment on GaAs bilayer \cite{khra} a profound reduction of the thermal activation gap has been found in some interval on the $\nu=2$ line.

In the case $\nu=1$ we parameterize general rotations from the denominator sub-group by four angles:
\begin{equation}\label{Fket}
|v_e\rangle=\left (\cos{\beta\over 2},\sin{\beta\over 2}\cos\alpha
e^{i\lambda_1},\sin{\beta\over 2}\sin\alpha e^{i\lambda_2}\right)
\end{equation} 
The skyrmion energy does not depend on the phases $\lambda_{1,2}$ and $\alpha=\pi/2$ in the energy minimum. We find Zeeman energy ${\cal Z}=h \left(1-\cos\beta\right)$, the valley splitting energy ${\cal T}=2t \cos^2(\beta/2)$ and the easy-plane Jahn-Teller energy ${\cal G}=G\delta \left(1+\cos\beta\right)$. The minimum of these skyrmion energies over the parameters $\beta$ and $\alpha$: ${\cal E}_{min}^{sk}=2\ {\rm min} \left(t+G\delta,h\right)$, determines the charge activation gap $\Delta$ in Eq.(\ref{GapSk}). The case $\nu=3$ is identical to the case $\nu=1$.

\section*{Conclusion}

We have shown that the valley degeneracy for Si(100) MOSFET can lead to Jahn Teller effect in strong magnetic field with a longitudinal lattice deformation inside the width of proper quantum well. This deformation has an atomic periodicity corresponding to the momentum difference between the two valley minima in the direction perpendicular to the plain. Integer fillings $\nu=1,3$ and $\nu=2$ are treated in different ways. For $\nu=1,3$ the Jahn Teller effect alone can not remove the valley degeneracy and Coulomb interaction is crucial. The level splitting here is small if the Coulomb energy is small compared to the Debay energy. For $\nu=2$ case Jahn-Teller effect removes the valley degeneracy via an easy-plane anisotropy similar to that found in the 2D bilayer system in strong magnetic field. Phase diagram depending on different physical parameters was established and anisotropic energy of topological skyrmion like texture was calculated. 

This work was supported by INTAS grant 99-01146 and RFBR grant 01-02-17520a. Author express their gratitude to V.M.Pudalov and V.T.Dolgopolov for usefull discussions.

\section*{Appendix A}

Here we calculate the second diagram on Fig.\ref{Diagrams} which was overlooked in Ref. \cite{IPF99}. It represents the change of Hartree-Fock exchange energy in a presence of spin texture. The spatial part of the electron Green function in Landau gauge is:
\begin{equation}\label{Gf} G_{0s}(z,z')=\frac{(z-z')^s}{\sqrt{2^ss!}}\ e^{- |z-z'|^2/4+i(x+x')(y-y')/2},
\end{equation}
where $z=x+iy$, $s$ is the Landau level and $G_{s0}=G_{0s}^*$. The bottom part of Fig.\ref{Diagrams}b features $G_{00}$ Green function whereas the upper part has $G_{10}$ and $G_{00}$ Green functions. Taking the two frequency integrals we find:
\begin{eqnarray}\label{e1} \delta E=\frac{1}{\sqrt{2}}\int \frac{d^2z}{2\pi}\frac{d^2z'}{2\pi}\frac{d^2\xi}{2\pi}\ G_{00}(z,z') V(z-z') \nonumber\\ \left(G_{00}(z'\xi)\Omega_+(\xi)G_{01}(\xi z)+G_{10}(z'\xi) \Omega_-(\xi) G_{00}(\xi z) \right),
\end{eqnarray}
Expanding $\Omega(\xi)$ around the center point of the diagram: $z_0=(z+z')/2$: $\Omega_\pm(\xi)= \Omega_\pm(z_0) \pm (\xi-z_0)\ \epsilon_{\mu\nu}\nabla_\mu\Omega^z_\nu(z_0)$, we evaluate the energy $\delta E$ (\ref{e1}) as ($z_0=\vec{r}$):
\begin{equation} \int d^2z\left(1-\frac{|z|^2}{4}\right)V(|z|)e^{- |z|^2/2}\int \epsilon_{\mu\nu}\nabla_\mu\Omega_\nu^z(\vec{r})\ \frac{d^2\vec{r}}{2\pi}
\end{equation}
The front integral is evaluated for the Coulomb interaction and equals: $(E_1+E_0)/2$.

\section*{Appendix B}

Eq.(\ref{Pi2Coset}) is prooved using a principal bundle of complex Stiffel manifold $V^C_{nk}=SU(n)/SU(n-k)$ - that is defined as a manifold of $k$ orthogonal complex vectors in $n$ dimensional complex linear space - over Grassmanian manifold $G^C_{nk}$ with a layer U($k$). The exact map sequence for this bundle is
\begin{equation} \label{exseq}
...\pi_2(V^C_{nk})\to \pi_2(G^C_{nk})\to \pi_1(U(k))\to \pi_1(V^C_{nk}) ...
\end{equation} 
Stiffel manifold has a property $\pi_j(V^C_{nk})=0$ at $j<2(n-k)$ \cite{Nov}, proven using the principal bundle SU($n$) over $V^C_{nk}$ with a layer SU($n-k$). Eq.(\ref{exseq}) means that $\pi_2(G^C_{nk})=\pi_1(U(k))$. The Lee group U($k$) is a product of SU($k$) and U(1) groups with the fundamental homotopy group: $\pi_1(U(k))=\pi_1(SU(k))+\pi_1(U(1))=Z$. \cite{Nov}

\end{document}